# Nonideal optical cavity structure of superconducting nanowire single photon detector

Hao Li, Weijun Zhang, Lixing You, Lu Zhang, Xiaoyan Yang, Xiaoyu Liu, Sijing Chen, Chaolin Lv, Wei Peng, Zhen Wang, and Xiaoming Xie

*Abstract*—Optical cavity structure has been proven to be a crucial factor for obtaining high detection efficiency in superconducting nanowire single photon detector (SNSPD). Practically, complicated fabrication processes may result in a non-ideal optical cavity structure. The cross-sectional transmission electron microscope (TEM) image of SNSPD fabricated in this study shows unexpected arc-shaped optical cavities which could have originated due to the over-etching of $SiO_2$ layer while defining NbN nanowire. The effects of the arc-shaped optical cavity structure, such as the wavelength dependence of the optical absorption efficiency for different polarization, were analyzed by performing optical simulations using finite-difference time-domain method. The central wavelength of the device is found to exhibit a blue shift owing to the arced cavity structure. This effect is equivalent to the flat cavity with a reduced height. The results may give interesting reference for SNSPD design and fabrication.

## I. Introduction

The increasing demand for fast, high detection efficiency and low dark count rate detectors has formed the genesis for the advancement of superconducting nanowire single-photon detectors (SNSPDs). Some of the potential or demonstrated applications of these devices include high data-rate lunar laser communications [1], quantum optics [2, 3], quantum information processing [4-6], noninvasive circuit testing [7], and depth imaging [8].

In principle, SNSPDs with high detection efficiency can be realized by solving the intricacies in the following three fundamental issues: optical coupling efficiency, absorption efficiency, and intrinsic detection efficiency [9, 10]. From the view point of optics, the absorption efficiency of the nanowire has to be increased as much as possible, and while maintaining the coupling loss as low as possible. Meanwhile, the dimension of the nanowire has to be approximately 5 nm (thickness) × 100 nm (width), so as to effectively respond to the infrared photons (i.e., high intrinsic detection efficiency). Given the condition, the width of the nanowire is one order of magnitude narrower than the diffraction limit of the incident infrared radiation. Consequently, the small overlap between the optical field and the nanowire mode severely limits the absorption efficiency of the nanowire. To this end, several studies have proposed various optical designs to enhance the absorption efficiency of the nanowire. For instance, the use of surface plasmon was expected to enhance the speed and efficiency of SNSPD by increasing the overlap of optical field[11, 12]. Besides, the fabrication of nanowire on an optical waveguide was also proposed to be an effective strategy, for increasing the nanowire-photon interaction length, which in turn would facilitate strong photon absorption [2, 13-15]. One of the most simple and popular methods, on-chip optical cavity structures, was also proven to be very effective to achieve high absorption efficiency, which virtually prolongs the nanowire-photon interaction length effectively [16-19]. Lately, studies have reported high system detection efficiency (SDE) in SNSPDs with the cavity structures reaching 93%/77%/76% for WSi /NbTiN/NbN respectively [18, 20, 21].

In the present study, we report the design and fabrication NbN SNSPD integrated with double cavity structures [10, 20] on double-side thermal oxidized silicon substrates. The cross-sectional microstructure of the fabricated SNSPD, as observed by using transmission electron microscope (TEM) indicated unexpected arc-shaped optical cavities. The origin of these non-ideal cavities could be attributed to the over-etching of the $SiO_2$ layer, while defining the NbN nanowire structure. Furthermore, we have numerically analyzed and discussed the influence of this non-ideal cavity structure on the absorption efficiency of SNSPD.

Hao Li, Weijun Zhang, Lixing You(Corresponding author), Lu Zhang, Xiaoyan Yang, Xiaoyu Liu, Sijing Chen, Chaolin Lv, Wei Peng, Zhen Wang and Xiaoming Xie are with the State Key Laboratory of Functional Materials for Informatics, Shanghai Institute of Microsystem and Information Technology, Chinese Academy of Sciences, Shanghai 200050, China (e-mail: lxyou@mail.sim.ac.cn) .

## II. experiments

In our experiment, 7-nm-thick NbN film was deposited onto double-side thermal oxidized silicon substrate using reactive magnetron sputtering in a gas mixture of Ar and $N_2$ (partial pressures of 90% and 10%, respectively). The thickness of $SiO_2$ layer in the substrate was 258nm, corresponding to the quarter wavelength for 1550nm. Subsequently, SNSPDs with a meander structure covering a square area of 15 μm × 15μm were fabricated from NbN films using e-beam lithography (EBL) and reactive ion etching (RIE). The linewidth and the period of the nanowire were 90 nm and 200 nm, respectively. A 200-nm-thick SiO layer was deposited on the top of the detector, which served as the dielectric material for the top cavity. Following that, a 150nm-thick silver film was deposited to create a mirror for the optical cavity. The bottom cavity is formed between the NbN layer and the silicon substrate, with $SiO_2$ as the dielectric material. When the photons are shined from the bottom side, the nanowire is located in between the two cavity structures, so as to ensure the high absorption efficiency. The thickness of each layer was designed to obtain the highest absorption efficiency at the wavelength of 1550 nm. Similar structure has been previously reported previously in the literature [9, 10, 16].

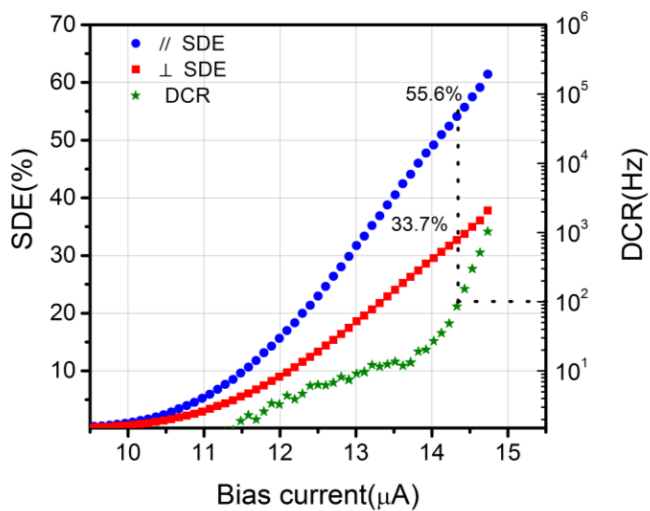

Fig. 1 System detection efficiency of 1550 nm for parallel and perpendicular polarization and DCR versus the bias current of SNSPD.

The SNSPD is mounted into a copper block packaging module, such that a lensed single mode fiber can be aligned directly from the backside of the substrate to the SNSPD [22]. The lensed fiber is specifically designed to ensure that the beam-waist of the light is smaller than the size of the meander structure, and its focus is located on the meandered nanowire. This packaging module ensures effective optical coupling between the fiber and device. The module is attached to the cold head of a two-stage Gifford–McMahon cryocooler with the working temperature of 2.20±0.005 K. For measuring the SDE, a continuous tunable laser is chosen as the photon source. The laser is heavily attenuated to obtain a photon flux of $10^6$ photons/s. Given the polarization sensitivity of the SNSPD, a parallel-polarized light with respect to the nanowire will have a maximal SDE, while the perpendicular-polarized light will result in a minimal SDE [19, 23]. A polarization controller was used to control the polarization of the incident photon. The SDE of the detector is defined as SDE = (PC − DCR)/P, where PC is the output pulse rate of the SNSPD, as measured by using a pulse counter; DCR is the dark count rate when the laser is powered off, and P is the photon rate input to the system. Fig. 1 shows the SDE at the wavelength of 1550 nm for both parallel and perpendicular polarization and the DCR as a function of the bias current for the SNSPD. The measured SDEs reach 55.6% and 33.7% at dark count of 100 Hz for the parallel and perpendicular polarization respectively.

It can be realized that the maximal SDE values are smaller than the corresponding theoretical values (shown in Fig. 4), implying that the SNSPD is non-ideal. The observed reduction in SDE could be ascribed to the imperfect optical coupling, non-ideal optical structure, and non-100% intrinsic detection efficiency. In the present study, we mainly focus on the absorptance of the optical structure. As mentioned earlier, the absorption efficiency is sensitive to the cavity structure formed by the film deposition process. Accordingly, the quality of the cavity structure was analyzed by using TEM, as shown Fig. 2. The cross-section of three nanowires in the center of the image shown in Fig. 2(a) indicates both the top cavity (Ag + SiO) and the bottom cavity ($SiO_2$ + Si) structures are registered clearly. However, the interface between Ag and SiO (see Fig. 2(b)) is found to exhibit an arc shape instead of a straight line. The height of the arc is approximately 17 nm, as measured from the TEM image. The magnified view of the nanowire area (see Fig. 2(c)) indicates that the $SiO_2$ around the nanowire was over-etched by 2 nm, which is necessary to ensure the quality of the NbN nanowire. Besides, the residue of the photoresist (>2 nm) was also observed on the NbN nanowire, implying the presence of a step with a height of over 10 nm on the edge of the NbN nanowire. As a SiO layer of thickness 200 nm is deposited on to the 10-nm-high step structure, it is reasonable to see an arc-shaped interface with the height of 17 nm between SiO and Ag layers.

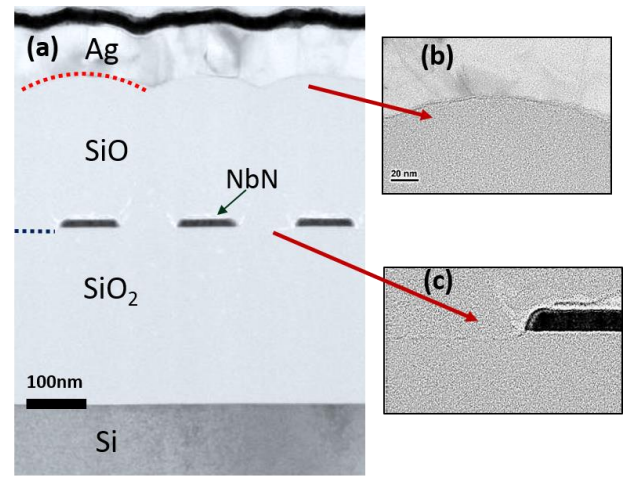

Fig. 2 TEM image of the SNSPD cross-section with the optical cavities. (a) Overall cross-sectional image showing three nanowires. The dotted line represents the interface of the different material layers. (b) Magnified view of the interface between the mirror Ag and SiO layers. (c) Magnified view of the nanowire revealing the over-etched $SiO_2$ and the residue of photoresist.

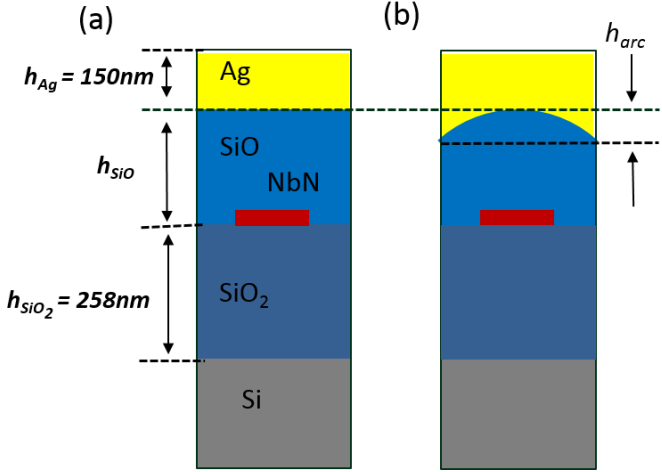

Fig. 3 (a) Ideal unit cell of the SNSPD. (b) Practical unit cell used in the numerical model of the absorption efficiency. Total height of the SiO layer is $h_{SiO}$ = 200 nm and the height of arc is 17 nm.

Furthermore, we gained deeper insight on the influence of the non-ideal optical cavity structure on the absorption efficiency of the SNSPD by performing numerical analysis using finite-difference time-domain (FDTD) method. Figure 3 shows the numerical model of the device. Figure 3(a) shows the designed ideal unit cell with SiO layer and $SiO_2$ layer of thickness of 200 nm and 258 nm, respectively, designed as a quarter of the wavelength. Figure 3 (b) shows the modeled unit cell with an arc shaped structure, obtained by considering the practical structure shown in Fig. 2. Here, we define the height of the arc as $h_{arc}$. Given the consideration that the height of the over-etched $SiO_2$ layer and the photoresist residue are just several nanometers, while the height of the arc of the cavity is of the order of tens of nanometers, the influence of the over etched $SiO_2$ layer and the photoresist residue on the absorption efficiency can be neglected in the calculation. The period of the unit cell is 200 nm and the width of the nanowire is 90 nm. For the calculations, we assumed that $n_{si}$ =3.45, $n_{sio2}$ =1.5, and $n_{sio}$=1.9, as measured by using spectroscopic ellipsometry. The optical properties of Ag and NbN layers were defined by using the Drude model. The corresponding refractive index at wavelength 1550nm are $n_{NbN}$= 5.23 + 5.82$i$, $n_{Ag}$= 0.51 + 10.71$i$ respectively[9, 24]. Periodic boundary conditions were applied for the left and right edges of the unit cell, and perfectly matched layer (PML) boundary conditions were applied on the bottom and top edges. This simulation is equivalent to an infinitely extended periodic structure in the horizontal direction, and therefore it neglects any edge effects of the real meander.

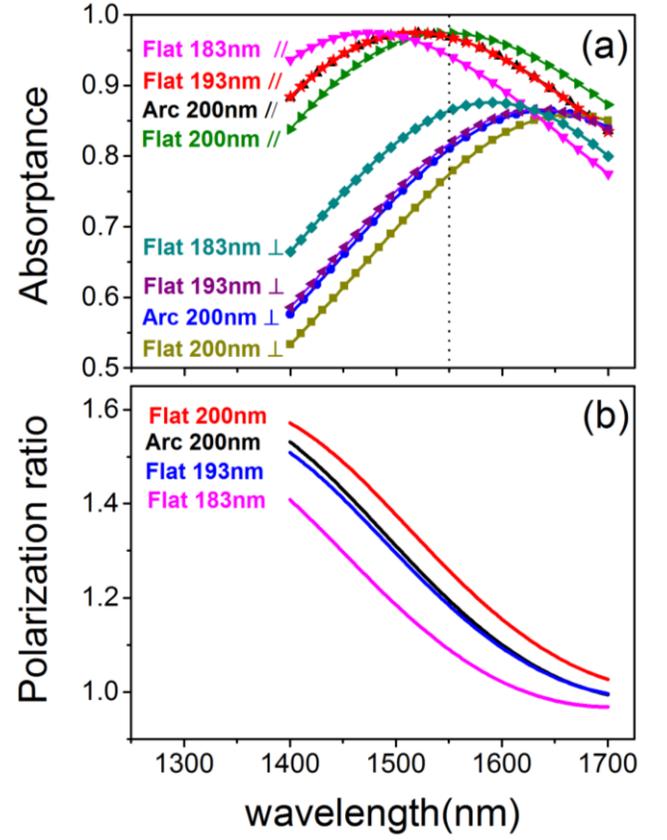

Fig. 4 (a) Calculated absorption efficiency for parallel and perpendicular polarizations versus wavelength for SNSPD with different parameters. The absorption efficiency for arc-shaped cavity structure with $h_{SiO}$ = 200 nm and $h_{arc}$ = 17 nm, flat cavity structure with $h_{SiO}$ = 183 nm, 193 nm and 200 nm, have been plotted. For parallel polarization, it can be observed that the absorption efficiency of the arc cavity structure almost overlaps with that of the flat cavity structure with cavity height of 193nm. On the other hand, the overlap is not so good for the perpendicular polarization. (b) Ratio of the absorption efficiency for different polarizations.

Figure 4 (a) shows the absorption efficiency plotted as a function of wavelength for the SNSPD with different parameters. The figure shows the absorption efficiency for the arc-shaped cavity structure with $h_{SiO}$ = 200 nm and $h_{arc}$ = 17 nm and that of the flat cavity structure with $h_{SiO}$ =183 nm, 193 nm, and 200 nm. The arc-shaped cavity structure has the same highest absorption efficiency for the parallel polarization. However, the center wavelength exhibits a blue shift with respect to the flat cavity structure. On the other hand, the arc-shaped optical cavity with $h_{SiO}$= 200 nm is equivalent to a flat optical cavity with $h_{SiO}$ = 193 nm. Similar results can also be found for the perpendicular polarization. However, we could observe a slight deviation in the absorption efficiency for the arc-shaped optical cavity with $h_{SiO}$ = 200 nm and the flat optical cavity with $h_{SiO}$ = 193 nm (see Fig 4(a)). The absorption ratio between the parallel polarization and perpendicular polarization for those structures was calculated, as shown in Fig. 4(b).

Here, the reduced thickness ($h_{off}$) is defined as the difference in thickness of $h_{SiO}$ for the arc-shaped cavity and the flat cavity, which has the same absorption efficiency characteristics. To further understand the effect of the arc-shaped cavity structure, we calculate the $h_{arc}$ dependence of $h_{off}$ for both parallel and perpendicular polarizations. As evidenced from Fig. 5, $h_{off}$ exhibit a linear relationship with $h_{arc}$ for both types of polarizations. It can be fitted by $h_{off}=0.59 \cdot h_{arc}$ and $h_{off}=0.73 \cdot h_{arc}$ for the parallel and the perpendicular polarizations, respectively. These results indicate that it may be necessary to optimize the thickness of SiO layer or control the fabrication process more precisely to obtain the highest absorption efficiency considering the possible influence contributed by the arc-shaped cavity structure.

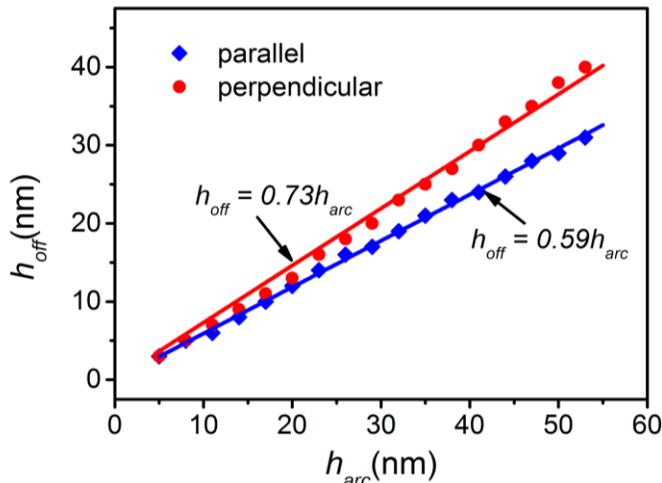

Fig. 5 Relation between $h_{off}$ and $h_{arc}$ for both the parallel and perpendicular polarizations.

III. CONCLUSION

In summary, high performance SNSPD with the double cavity structure was fabricated. The arc-shaped cavity structure was observed in the cross-sectional TEM image, which could be attributed to the over-etching of the $SiO_2$ layer, while defining the nanowire structure. Numerical simulations indicate that the arc-shaped cavity structure is equivalent to a flat cavity structure with a reduced thickness for the parallel polarization. Besides, we also determined the arc height dependence of the reduced thickness. The results obtained in this study provide interesting information on obtaining high absorption efficiency by optimizing the fabrication parameters.

ACKNOWLEDGMENT

This work was supported in part by the National Natural Science Foundation of China (Grant 91121022), 973 Program (Grant 2011CBA00202), 863 Program (Grant 2011AA010802) and in part by the "Strategic Priority Research Program (B)" of the Chinese Academy of Sciences (Grant XDB04010200).

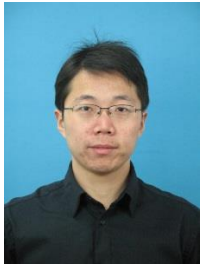

**Hao Li** received his B.S. degree from Jilin University in July 2008, and obtained his Ph.D. degree from Shanghai Institute of Microsystem and Information Technology, Chinese Academy of Sciences, in 2013. Currently, his research interests focus on the design and experimental characterization of superconducting single photon optical detectors

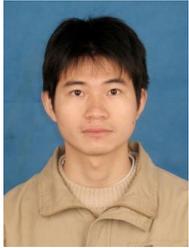

**Weijun Zhang** received his Ph.D. degree in condensed matter physics from the Institute of Physics, Chinese Academy of Sciences (CAS), Beijing, China, in 2012. He is currently an associate researcher with the SNSPD Group, State Key Laboratory of Functional Materials for Informatics, Shanghai Institute of Microsystem and Information Technology (SIMIT), CAS, Shanghai. His current research interests include electrical and magnetic properties of nanostructures, superconducting devices and physics, and single-photon detectors.

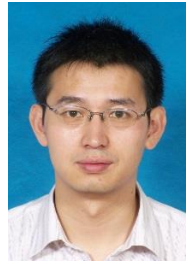

**Lixing You** received his B.S., M.S., and Ph.D. degrees in physics from Nanjing University, Nanjing, China, in 1997, 2001, and 2003, respectively. From 2003 to 2005, he worked as a Postdoctoral Researcher in the Department of Microtechnology and Nanoscience (MC2), Chalmers University of Technology, Göteborg, Sweden. Later on, from 2005 to 2006, he worked as a Postdoctoral Researcher with the Condensed Matter Physics and Devices Group, University of Twente, Twente, The Netherlands. Since 2006, he has been a Guest Researcher with the Electromagnetics Division, National Institute of Standards and Technology, Boulder, CO. Since September 2007, he has been a Research Professor with Shanghai Institute of Microsystem and Information Technology, Chinese Academy of Sciences, Beijing, China. His research focuses on superconductive electronics, including micro/nano superconductive devices and high-frequency applications.

**Lu Zhang** received her B.S. degree from Yangtze University in July 2008, and M.S. degree in condensed matter physics from Soochow University, in 2011. Her current research interest lies in the deposition and characterization of thin films.

**Xiaoyan Yang** received her B.S. degree from University of electronics science and technology, Chengdu, China, in 2005, and M.S. degree from Shanghai University, Shanghai, China, in 2008. Currently, she is pursuing her Ph.D. degree at the Shanghai Institute of Microsystem and Information Technology, Chinese Academy of Sciences.

**Xiaoyu Liu** received his M.S. degree from Nanjing University, in July 2011. Currently, he is working on the Electron Beam Lithography process.

**Sijing Chen** received his B.S. degree from Huazhong University of Science and Technology, in July 2008, and obtained his Ph.D. degree from Shanghai Institute of Microsystem and Information Technology, Chinese Academy of Sciences, in 2013. His current research activities include the experimental characterization of superconducting single photon optical detectors.

**Chaolin Lv** received his B.S. degree from Jilin University, in July 2013. Currently, he is working towards his Ph.D. degree at the Shanghai Institute of Microsystem and Information Technology, Chinese Academy of Sciences.

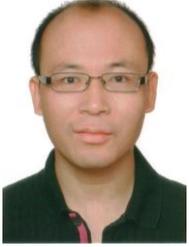

**Wei Peng** received his B.S. and M.S. degrees from the School of Physics and Electronic Technology, Hubei University, in 1997 and 2000, respectively, and Ph.D. degree in Condensed Matter Physics from the Institute of Physics, Chinese Academy of Sciences, in 2004. From 2004 to 2010, he was actively engaged in research at Tsinghua University (China), Université de Rennes1 (France), Université de Picardie Jules Verne (France), and Max-planck Institute für Mikrostrukturphysik (Germany). Since November 2010, he has been a Professor at the Shanghai Institute of Microsystem and Information Technology, Chinese Academy of Sciences. His research interests include micro- and nanofabrication technologies for superconductive devices.

**Zhen Wang** obtained his Ph.D. degree in electrical engineering degree from Nagaoka University of Technology, Nagaoka, Japan, in 1991. From 1991 to 2013, he worked at the National Institute of Information and Communications Technology (NICT), Japan. He is currently a Research Professor at the Shanghai Institute of Microsystem and Information Technology (SIMIT), Chinese Academy of Science, China. He is Fellow of NICT. His research interests focus on superconducting electronics, including superconducting devices and physics, superconducting SIS terahertz mixers, and photon detectors.

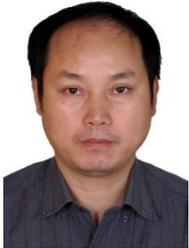

**Xiaoming Xie** obtained his Ph.D. degree in 1990 from the Shanghai Institute of Microsystem and Information Technology, Chinese Academy of Sciences. Between April 1993 and Nov. 1995, he worked as a postdoctoral research in the group of "Surfaces et Supraconducteurs" at the Ecole Supérieure de Physique et de Chimie Industrielles de la Ville de Paris, France. From Nov. 1995 to May 2005, he conducted extensive research on electronic materials and reliability of electronics packaging. Since May 2005, he has been mainly focusing on SQUID applications. His major research interests encompass quantum materials, superconducting sensors and detectors, and their applications.